\documentstyle[12pt,epsf]{article}
\begin{document}
\textwidth 13.6cm
\textheight 19.9cm
\noindent {\large \bf Equilibrium statistics of an infinitely long chain in the severe confined geometry: Rigorous results}

\noindent {\it \large Pramod Kumar Mishra} \\
\noindent {\bf Department of Physics, DSB Campus, Kumaun University \\ Nainital-263 002, India} \\

\noindent {\bf Abstract} :
We analyze the equilibrium statistics of a long linear  homo-polymer chain confined 
in between two flat geometrical constraints under good solvent condition. 
The chain is occupying two dimensional space and geometrical constraints are two impenetrable lines for the two dimensional space. 
A fully directed self avoiding walk lattice model is used to derive analytical expression of the partition function for the
given value of separation in between the impenetrable lines. The exact values of the critical exponents
($\nu_{||}, \nu_{\perp}, \nu $ and $ \gamma_1$) were obtained for different value of separations in between the impenetrable lines. An
exact expression of the grand canonical partition function of the confined semiflexible chain is also calculated for the given value of the
constraints separation using generating function technique.

\vspace {.2cm}
\section{Introduction:}
The subject of confinement of a long linear polymer chain has received much attention in the past a few years due to advances
in single molecule based experiments \cite{1,2,3,4,5,6}, and also see references quoted therein. 
It is now possible to measure conformational properties of a confined polymer chain \cite{7}. Such study
can reveal wealth of useful information regarding elastic properties and molecular organization of the chain in the crowded geometry. It may
also be useful in understanding molecular processes occurring in the living cells. 

Lattice model of self-avoiding walk is extensively used to study conformational properties of a long linear polymer chain. 
A rich phase diagram of the surface interacting polymer chain has been reported by various authors \cite{1,2,3,4,5,6,7,8}. 
The values of critical exponents were also reported and these exponents were found useful in understanding thermodynamics of the chain. 
However, a few aspects of a confined long chain in the severe confined geometry still 
require attention. These studies may find useful applications in the steric stablisation of the colloids dispersions, 
surface coating and adsorption behaviour of the gels \cite{1,2,3,4}. 

In this paper we consider a long chain confined in between two impenetrable lines. The separation in between confining 
lines is varied in the unit of monomer size. An approach of canonical ensemble theory is used to derive exact expression of the canonical partition
function of the confined flexible chain for a given value of separation 
($M$) in-between the confining lines. The value of critical exponents $\nu_{||}, \nu_{\perp}, \nu$ and
$\gamma_1$ were determined for the separation ($M$) in between the confining lines. We have also derived the exact expression of the grand
canonical partition function of the confined semi-flexible chain using generating function technique.

The paper is organized as follows: In Sec. 2, lattice model of
fully directed self avoiding walk is described for a confined linear polymer chain 
under good solvent condition on a square lattice. In sub-section 2.1, we use canonical ensemble theory to
derive expression of the thermodynamic parameters of the confined single chain. Analytical expressions of the 
critical exponents were obtained for the confined flexible chain. In the subsection 2.2 the grand canonical ensemble 
theory is used to derive exact partition function of an infinitely long semiflexible chain for a given
value of separation in between the confining impenetrable lines.  
Finally, in Sec. 3 we summarize and discuss the results obtained.

\section{Model and method}
A model of fully directed self-avoiding walks \cite{9} on a square lattice is used to 
investigate the possibility of an adsorption transition of an infinitely long linear semiflexible homopolymer chain 
on geometrical constraints, when the chain is confined in between two
impenetrable stair shaped surfaces under good solvent condition (as shown schematically in figure 1).  The directed walk model is
restrictive in the sense that the angle of bending has a unique value, that is $90^{\circ}$ for a square lattice
and directedness of the walk amounts to certain degree of stiffness in the 
walks of the chain because the different
directions of the space are not treated equally. 
Since, the directed self avoiding walk model can be solved 
analytically and therefore gives exact values of the  
partition function of the polymer chain.  
We consider a fully directed self avoiding walk ($FDSAW$) 
model, therefore, the walker is allowed to take steps along $+x$, and $+y$ directions on a square lattice in between the constraints.

\begin{figure}[htbp] 
\centering 
\epsfxsize=8cm\epsfbox{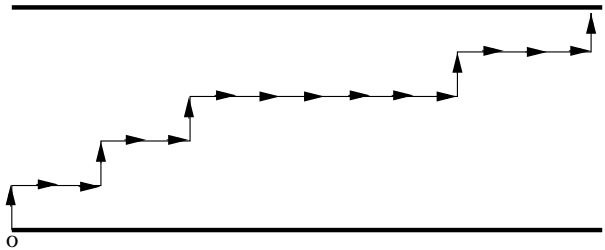} 
\caption{
This schematic diagram shows a linear polymer chain confined in between two impenetrable lines. The lines are $y=0$ and $y=5$ {\it i. e.}
the separation ($M$) in between the confining lines is 5 units. One end of the confined chain is grafted at ($O$) lower impenetrable line. 
}
\label{Figure1}
\end{figure}

A fully directed self avoiding walk $\cite{1}$ is used to
model a long linear homo-polymer chain in between two impenetrable lines ($y=0$ and $y=M$) on a square lattice under 
good solvent condition (as shown in figure-1, where $M=5$). In other words, the separation is five units in-between impenetrable lines. 
One end of the confined chain is grafted on an impenetrable line $y=0$ and 
the walker is allowed to take steps along $+x$ and $+y$ directions in between the two impenetrable lines. 
The spacing in-between two impenetrable lines is increased by one unit of monomer (step) size. 
Since, the walk is fully directed; therefore, it allows certain degree of restriction on the chain. However, directed
walk model is analytically solvable \cite{1,3,6,8,9,10} and exact values of the critical exponents can be determined using this model.

The stiffness of the chain is accounted by introducing an energy $\epsilon_b(>0)$ for each bend occurring in the chain. 
The stiffness weight is $k=exp(-\beta\epsilon_{b})$, where $\beta=(k_BT)^{-1}$ is 
inverse of the temperature. If, $k=1$ or $\epsilon_{b}=0$, the chain is flexible and for 
$0<k<1$ or $0<\epsilon_{b} <\infty$ the chain is  
semiflexible. However, if $\epsilon_{b}\to\infty$ or $k\to0$, 
the polymer chain is like a rigid rod.

The generating function technique is
used to derive exact expression of the partition function of a confined semiflexible chain, when the separation in between two
impenetrable lines is $M$ units.

\subsection{A long linear chain confined in between the impenetrable lines: Canonical Ensemble Approach}

We consider a long linear chain of $N$ monomers confined in between two impenetrable lines. The lines are separated to each other by
a distance of $M$ units. The number of conformations [$C_N^M(N_x,N_y)$] of the confined chain varies with number of steps $N_x$ and $ N_y$
(where, $N_x$ is the number of steps along $+x$ direction and $N_y$ is the number of steps along $+y$ direction) of a walk of
$N$ steps in between the confining lines. A list of values of the conformations are shown in the table -1 of a confined flexible chain 
of $N$ monomers and the separation in between the confining impenetrable lines is $M$.

The total number of conformations $[Z_N^M]$ of the confined chain of $N$ monomers variaes with separation $M$ in between the impenetrable lines
and its value for a flexible chain is written as,

\begin{equation}
Z_N^M=\sum_{N_y=0}^MC_N^M(N_x,N_y)= \sum_{K=1}^M\frac{\Pi_{P=1}^K[N-(P-1)]}{K!}+1
\end{equation}

However, the number of suppressed conformations of the confined flexible chain of $N$ monomers is $2^N-Z_N^M$, when the separation is $M$ units
in between the confining impenetrable lines.
   
An average value of the parallel component ($R_x(M)$) of the end-to-end distance ($R_E(M)$) is calculated using following relation,

\begin{equation}
<R_x^2(M)>=\frac{ \sum_{N_y=0}^M (N-N_y)^2C_N^M(N_x,N_y)}{Z_N^M}
\end{equation}
where, $R_x(M)=N_x*a=(N-N_y)*a$, $a$ is lattice parameter of a square lattice. The value of $a$ is taken unity for the sake of 
mathematical simplicity.

In the long length limit of the confined flexible chain $R_x(M)$ scales as,
\begin{equation}
<R_x(M)>_{N\to\infty}\sim N(1+\frac{M}{N})
\end{equation}
While, the average value of the perpendicular component $R_y(M)$ of the end-to-end distance of the confined chain is determined 
by following equation,


\begin{table}

\caption{The number of conformations ($C_N^M(N_x,N_y)$) of a N step long confined flexible chain 
is shown in this table. The separation is $M$ units in between the confining impenetrable lines.}

$N_x$\hspace{2.9cm}$N_y(=N-N_x)$\hspace{2.5cm}$C_N^M(N_x,N_y)$\\
\hspace{1cm}N-M\hspace{2.9cm}M\hspace{3cm}$\frac{N(N-1)(N-2)\dots [N-(M-1)]]}{M!}$\\
\hspace{1cm}N-(M-1)\hspace{2cm}M-1\hspace{3cm}$\frac{N(N-1)(N-2)\dots [N-(M-2)]]}{(M-1)!}$\\
\hspace{1cm}N-(M-2)\hspace{2cm}M-2\hspace{3cm}$\frac{N(N-1)(N-2)\dots [N-(M-3)]]}{(M-2)!}$\\
\hspace{1cm}N-(M-3)\hspace{2cm}M-3\hspace{3cm}$\frac{N(N-1)(N-2)\dots [N-(M-4)]]}{(M-3)!}$\\
\hspace{1cm}.\hspace{4cm}.\hspace{3cm}.\\
\hspace{1cm}.\hspace{4cm}.\hspace{3cm}.\\
\hspace{1cm}.\hspace{4cm}.\hspace{3cm}.\\
\hspace{1cm}.\hspace{4cm}.\hspace{3cm}.\\
\hspace{1cm}.\hspace{4cm}.\hspace{3cm}.\\
\hspace{1cm}N-4\hspace{3cm}4\hspace{3cm}$\frac{N(N-1)(N-2)(N-3)}{4!}$ \\
\hspace{1cm}N-3\hspace{3cm}3\hspace{3cm}$\frac{N(N-1)(N-2)}{3!}$ \\
\hspace{1cm}N-2\hspace{3cm}2\hspace{3cm}$\frac{N(N-1)}{2!}$ \\
\hspace{1cm}N-1\hspace{3cm}1\hspace{3cm}N \\
\hspace{1cm}N\hspace{3.4cm}0\hspace{2.0cm}\hspace{1cm}1\\

\end{table}

\begin{equation}
<R_y^2(M)>=\frac{ \sum_{N_y=0}^M N_y^2C_N^M(N_x,N_y)}{Z_N^M}
\end{equation}
where, $R_y(M)= N_y*a$.

In the long length limit of the confined flexible chain $R_y(M)$ scales as,
\begin{equation}
<R_y(M)>_{N\to\infty}\sim M(1+\frac{M}{N})
\end{equation}

The average value of end-to-end distance is,

\begin{equation}
<R_E^2(M)>=\frac{ \sum_{N_y=0}^M [N_y^2+(N-N_y)^2]C_N^M(N_x,N_y)}{Z_N^M}
\end{equation}
The universal configurational exponent $\gamma_1$ of the confined flexible chain with one end grafted on an impenetrable line ($y=0$)
is calculated using formula,

\begin{equation}
\gamma_1(M)=\frac{Ln(\frac{Z_{N+P}^MZ_{N-P}^M}{(Z_N^M)^2})}{Ln(\frac{N^2-P^2}{N^2})}
\end{equation}
where, $N$ and $P$ is the number of monomers in the confined flexible chain.
We have found in the long length limit of the confined chain,
\begin{equation}
\frac{Z_{N+P}^{M}Z_{N-P}^{M}}{(Z_N^{M})^2}=\frac{N^2-P^2}{N^2}
\end{equation}
The value of $\gamma_1(M)$=1.
\begin{figure}[htbp] 
\centering 
\epsfxsize=8cm\epsfbox{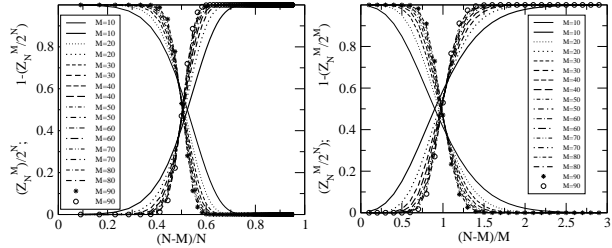} 
\caption{
It has been shown in this figure that the impenetrable lines (geometrical constraints) supress huge number of 
conformations of the confined flexible chain. When $N=2M$, fifty percent conformations 
of the confined flexible chain is not polymerized due to presence of impenetrable lines.}
\label{Figure1}
\end{figure}

\subsection{\bf \it An infinitely long semiflexible chain in between two impenetrable lines: Grand Canonical Ensemble approach}
The grand canonical partition function of an infinitely long linear semiflexible homo-polymer chain confined in between two
impenetrable lines (as shown in figure-1) is written as, 

\begin{equation}
G(k)={\sum}^{N=\infty}_{N=0}\sum_{ all\hspace{0.07cm}walks\hspace{0.07cm}of\hspace{0.05cm}N\hspace{0.05cm}steps} {g}^{N}k^{N_b}
\end{equation}
where, $N_b$ is the total number of bends in a self avoiding walk of $N$ steps (bonds or monomers),
and $g$ is step fugacity of the walker making steps in the space lying in-between the impenetrable lines.

The method of generating function technique $\cite{1}$ is used to calculate the grand canonical 
partition function of the confined semiflexible homopolymer chain (as shown in figure-1).

The components [$X_M(k,g) \& Y_M(k,g)$] of the grand canonical partition function $G_M(k,g)$ of the confined semiflexible chain,
when the separation in-between the confining  impenetrable lines is equal to $M$ units, can be written as,

\begin{equation}
X_M(k,g)=\frac{g}{1-g}+\frac{gk(1-g+gk)}{(1-g)^2} [u_1+u_2+u_3+u_4+u_5] 
\end{equation}

where, 
$u_1= \sum_{P=1}^Mg^P+\frac{g^2k^2}{1-g}\sum_{P=1}^{M-1}Pg^P$ 

$u_2=[\frac{g^2k^2}{1-g}]^2\sum_{P=1}^{M-2} \frac{P(P+1)}{2!}g^P +[\frac{g^2k^2}{1-g}]^3\sum_{P=1}^{M-3}\frac{P(P+1)(P+2)}{3!}g^P + \dots $

$u_3=[\frac{g^2k^2}{1-g}]^{M-3}\sum_{P=1}^3\frac{P(P+1)(P+2)\dots (P+M-4)}{(M-3)!}g^P+ $

$u_4=[\frac{g^2k^2}{1-g}]^{M-2}\sum_{M=1}^2\frac{P(P+1)(P+2)+\dots (P+M-3)}{(M-2)!}g^P+ $

$u_5=[\frac{g^2k^2}{1-g}]^{M-1}\sum_{P=1}^1\frac{P(P+1)(P+2)\dots(P+M-2)}{(M-1)!}g^P$

and $P \le M$ (and $M$=1,2,3,4...,$\infty$) and $X_M(k,g)$ is the sum of the Boltzmann weight of all the walks having 
first step along $+x$-direction. 

\begin{equation}
X_{M\to\infty}(k,g)=\frac{g}{1-g-gk}
\end{equation}
provided $g<\frac{1}{1+k}$ {\it i. e.} $g_c$ \cite{8}.

Similarly, $Y_M(k,g)$ is the sum of Boltzmann weight of all the walks of the confined polymer 
chain having first step along $+y$ direction. It can be written as,

\begin{equation}
Y_M(k,g)=\frac{1-g+gk}{1-g} [v_1+v_2+v_3+v_4+v_5]
\end{equation}
where,

$v_1=\sum_{P=1}^Mg^P+\frac{g^2k^2}{1-g}\sum_{P=1}^{M-1}Pg^P+ $ 

$v_2=[\frac{g^2k^2}{1-g}]^2\sum_{P=1}^{M-2} \frac{P(P+1)}{2!}g^P +[\frac{g^2k^2}{1-g}]^3\sum_{P=1}^{M-3}\frac{P(P+1)(P+2)}{3!}g^P + \dots $

$v_3=[\frac{g^2k^2}{1-g}]^{M-3}\sum_{P=1}^3\frac{P(P+1)(P+2)\dots (P+M-4)}{(M-3)!}g^P+ $

$v_4=[\frac{g^2k^2}{1-g}]^{M-2}\sum_{M=1}^2\frac{P(P+1)(P+2)+\dots (P+M-3)}{(M-2)!}g^P+ $

$v_5=[\frac{g^2k^2}{1-g}]^{M-1}\sum_{P=1}^1\frac{P(P+1)(P+2)\dots(P+M-2)}{(M-1)!}g^P]$

and,
\begin{equation}
Y_{M\to\infty}(k,g)=\frac{g}{1-g-gk}
\end{equation}
for $g<g_c$.
The grand canonical partition function ($Z_M(k,g)$) of the confined chain can be calculated by summing the above two components of the partition function,

\begin{equation}
Z_M(k,g)=\frac{g}{1-g}+[\frac{1-g+gk}{1-g}]^2 [w_1+w_2+w_3+w_4+w_5]
\end{equation}

where,

$w_1=\sum_{P=1}^Mg^P+\frac{g^2k^2}{1-g}\sum_{P=1}^{M-1}Pg^P+ $

$w_2=[\frac{g^2k^2}{1-g}]^2\sum_{P=1}^{M-2} \frac{P(P+1)}{2!}g^P +[\frac{g^2k^2}{1-g}]^3\sum_{P=1}^{M-3}\frac{P(P+1)(P+2)}{3!}g^P + \dots $

$w_3=[\frac{g^2k^2}{1-g}]^{M-3}\sum_{P=1}^3\frac{P(P+1)(P+2)\dots (P+M-4)}{(M-3)!}g^P+ $

$w_4=[\frac{g^2k^2}{1-g}]^{M-2}\sum_{M=1}^2\frac{P(P+1)(P+2)+\dots (P+M-3)}{(M-2)!}g^P+ $

$w_5=[\frac{g^2k^2}{1-g}]^{M-1}\sum_{P=1}^1\frac{P(P+1)(P+2)\dots(P+M-2)}{(M-1)!}g^P] $

and substituting $M\to\infty$ in above equation of the partition function, 
we recover $Z_\infty(k,g)=\frac{2g}{1-g-gk}$ \cite{8}, provided $g<g_c(=\frac{1}{1+k})$ \cite{8}.

\section{Result and discussion}
A fully directed self avoiding walk lattice model is used to calculate canonical partition function of a long linear 
flexible polymer chain confined in between two impenetrable lines (as shown in figure-1). The confined chain is uccupying space
in between two confining lines and the chain is under good solvent condition. The separation ($M$) in-between the confining impenetrable 
lines is increased in the unit of monomer (step) size. The exact expression of the canonical partition function of the confined flexible
chain is obtained for a given value of the separation in between the confining lines. The critical value of exponents were determined and 
the exact expression of the grand canonical partition function of a confined semiflexible chain is also derived. 
We have chosen monomer (step) size unity for the sake of mathematical simplicity. 

We have shown the variation in the number ($C_N^M(N_x,N_y)$) of conformations (entropy) of the confined flexible chain 
of $N$ monomers for 
a given separation ($M$) in between the confining lines (as shown in table-1).  
The effect of confinement on the statistics of a flexible chain is illustrated by the figure 2. 
It has been found that the
50$\%$ conformations of the confined flexible chain are suppressed due to confinement, when $N=2M$. The number of suppressed conformations 
is $2^N-Z_N^M$  of a confined flexible chain of $N$ monomers confined in between two impenetrable lines separated by $M$ units.

The average value of end-to-end distance of the confined flexible polymer chain is modified due to confinement. 
The average value of the parallel component ($R_x(M)$) of the end-to-end distance $[R_E(M)]$ and $R_E(M)$ 
of the confined flexible chain scales as, $<R_x>\sim N(1+\frac{M}{N})$ and the perpendicular component of end-to-end distance ($R_y(M)$)
scales with the confined chain size as $<R_y>\sim M(1+\frac{M}{N})$. Therefore, the value of $\nu_{||}=1$ and $\nu=1$ while, 
$\nu_{\perp}=0$ for the confined flexible chain, provided the value of $M$ is finite. 
It has been concluded from the value of critical exponent $\nu$ and $\nu_{||}$
that the confined chain in quasi-one dimension. Therefore, value of $\gamma_1(M)$ is unity for a flexible
chain confined in-between two parallel impenetrable lines.  

It has also been found that the critical value of step fugacity ($g_c$) of the confined 
flexible chain is 1, when the separation in-between the confining impenetrable lines is $M$. However,
$g_c=\frac{1}{1+k}$\cite{8} as the value of $M\to\infty$.

\subsection{Acknowledgements}
Author thankfully acknowledge financial support received from Department of Science and Technology, New Delhi (Project no. SR/FTP/PS-122/2010). 

\end{document}